\documentstyle[12pt]{article}
\newcommand{\sbody}[2]{{\textstyle\frac{#1}{#2}}}

\begin{document}
\begin{center}
\vfill
\large\bf{Operator Regularization and}\\
\large\bf{Noncommutative Chern Simons Theory}
\end{center}
\vfill
\begin{center}
D.G.C. McKeon\\
Department of Applied Mathematics\\
University of Western Ontario\\
London\\
CANADA\\
N6A 5B7
\end{center}
\vspace{.1cm}
Tel: 519-661-2111, ext. 88789\\
Fax: 519-661-3523\\
Email: DGMCKEO2@UWO.CA
\vspace{.3cm}
\section{Abstract}

We examine noncommutative Chern Simons theory using operator regularization. 
Both the $\zeta$-function and the $\eta$-function are needed to
determine one loop effects. The contributions to these functions coming
from the two point function is evaluated. The $U(N)$ noncommutative
model smoothly reduces to the $SU(N)$ commutative model as the
noncommutative parameter $\theta_{\mu\nu}$ vanishes.

\section{Introduction}

Normally, Chern Simons theory is a purely topological theory as it is
metric independent [1-3]. However, if we consider Chern Simons theory in
a noncommutative space [4,5] in which
$$\left[x_\mu , x_\nu \right] = -i\theta_{\mu\nu}\eqno(1)$$
metric dependence inevitably arise. The consequences of the presence of
$\theta_{\mu\nu}$ in noncommutative Chern Simons theory have been
considered in [6].  In particular, the one loop contribution to the two
point function has been computed in [7].

In this paper, we would like to examine this one loop, two point
function using operator regularization [8], a generalization of 
$\zeta$-function 
regularization [9]. The modulus and phase of the functional determinant associated with the
effective action at one loop order are associated with the $\zeta$-function [8,9]
and $\eta$-function [1, 10-12] respectively.  This approach has been
used in conventional Chern Simons theory [11-15].

\section{The $\zeta$- and $\eta$-functions}

The usual Chern Simons action is given by
$$S = \int d^3x \,Tr\, \epsilon_{ijk} \left( A_i\partial_j A_k - \frac{2i}{3}
A_i A_j A_k\right)\eqno(2)$$
where $A_i = A_i^a T^a$ with $T^a$ being the generator of a
Lie group $G$.  If $G$ is a unitary group $U(N)$, then (2) can be generalized
by converting the products occuring in (2) to Moyal product in which [4,5]
$$\left. A_i(x) * A_j(x) = e^{-\frac{i}{2}\partial^x \times \partial^y} A_i(x) A_j(y)
\right|_{x=y}\eqno(3)$$
where $a \times b = \theta_{ij} a_i b_j$.

In order to use operator regularization to compute radiative effects,
it is necessary to employ background field quantization [16-18].  We begin
by splitting $A_i$ into the sum of background and quantum fields
$$A_i \rightarrow A_i + Q_i .\eqno(4)$$
A gauge fixing Lagrangian
$$L_{gf} = Tr \left(N D_i (A) Q_i \right)\eqno(5)$$
is chosen so as to leave the symmetry
$$A_i \rightarrow A_i + D_i (A)\Omega\eqno(6a)$$
$$Q_i \rightarrow Q_i - i\left(Q_i * \Omega - \Omega * Q_i\right)\eqno(6b)$$
$$ \equiv Q_i - i \left[Q_i , \Omega\right]\nonumber$$
present in (2) unbroken.  In (5), $N$ is a Nakanishi-Lautraup field and $D_i(A)$
is the covariant derivative
$$D_i (A) f = \partial_i f - i \left[ A_i , f\right].  \eqno(7)$$
The ghost Lagrangian associated with the gauge fixing of eq. (5) and the gauge transformation
$$A_i \rightarrow A_i + D_i (A + Q)\Lambda \eqno(8a)$$
$$Q_i \rightarrow Q_i\eqno(8b)$$
is
$$L_{\rm{ghost}} = \overline{C} D_i(A) D_i(A + Q)C.\eqno(9)$$
From (2), (5) and (9) it is evident that the terms in the effective Lagrangian
$L + L_{gf} + L_{\rm{ghost}}$ that are bilinear in the quantum fields are
$$ L^{(2)} = \left(Q_i^a , N^a\right) \left(\begin{array}{cc}
\epsilon_{ipj} D_p^{ab}(A) & -D^{ab}_i (A)\\
D_j^{ab}(A) & 0 \end{array}\right)\left(\begin{array}{c}
Q_j^b \\
N^b\end{array}\right)\nonumber$$
$$+ \overline{C}^a \left(D_i^{ab} (A) D_i^{bc} (A) \right)C^c ,\eqno(10)$$
so that the one loop generating functional is given by
$$W^{(1)} (A) = \left( \ln {\rm{det}}^{-1/2} H_I\right) + \left(\ln \det H_{I\!I}\right)\eqno(11)$$
where $H_I$ and $H_{I\!I}$ are the two operators in (10).  As $H_I$ is linear in derivatives,
it is necessary to make the replacement
$$\ln{\rm{det}}^{-1/2} H_I \rightarrow \ln{\rm{det}}^{-1/4} H_I^2 .\eqno(12)$$
(A possible loss of phase in this replacement will be considered below.) Since
$$\epsilon_{ijk} D_j(A) D_k (A) f = -\frac{i}{2} \epsilon_{ijk}\left[F_{jk} ,
f\right]\eqno(13)$$
where
$$F_{jk} = \partial_i A_j - \partial_jA_i - i\left(A_i *
A_j - A_j * A_i\right), \eqno(14)$$
it is evident that
$$H_I^2 = \left(\begin{array}{cc} -D^2\delta_{ij} & -\epsilon_{imn} F_{mn}\\
\epsilon_{mnj} F_{mn} & -D^2\end{array}\right).\eqno(15)$$
We note that
$$[M,N] = M^a T^a * N^b T^b - N^b T^b * M^a T^a\nonumber$$
$$=\frac{1}{2} \left( \left[ T^a, T^b\right] \left\lbrace M^a, N^b \right\rbrace
\right.\eqno(16)$$
$$\left. + \left\lbrace T^a, T^b \right\rbrace \left[ M^a, N^b \right]\right)\nonumber$$
$$\equiv \left(if^{abc} \left\lbrace M^a, N^b \right\rbrace + d^{abc}
\left[M^a, N^b \right]\right)T^c\nonumber$$
and this becomes, on account of the Moyal product defined in (3)
$$ = 2i\left[ f^{abc} \cos \left( \frac{p \times q}{2}\right) + d^{abc}
\sin\left(\frac{p \times q}{2}\right)\right] (M^aN^bT^c)\eqno(17)$$
where $p$ and $q$ are the momenta of $M^a$ and $N^b$ respectively.
This is to be utilized when employing a perturbative expansion of $W^{(1)}(A)$
in powers of $A$. (Conventions for $f^{abc}$ and $d^{abc}$ are those of [19].)

In operator regularization, one first uses
$$\hspace{-3cm}\ln det H = tr\ln H\nonumber$$
$$= \lim_{s\rightarrow 0} - \frac{d}{ds} \left(tr H^{-s}\right)\nonumber$$
$$\left. = - \frac{d}{ds}\right|_0 \frac{1}{\Gamma(s)} \int_0^\infty dt\,t^{s-1}
tr e^{-Ht} \eqno(18)$$
$$\equiv - \zeta^\prime (0)\nonumber$$
and then employs an expansion due to Schwinger [20]
$$tr e^{-\left(H_0 + H_1\right)t} = tr\left[ e^{-H_0t} + \frac{(-t)}{1} e^{-H_0 t} H_1
\right. \nonumber$$
$$\left. + \frac{(-t)^2}{2} \int_0^1 du\,e^{-(1-u)H_0 t} H_1 e^{-uH_0 t} H_1 + \ldots 
\right]\eqno(19)$$
to effect an expansion in powers of the background field. (The dependence of $H$ on the
background field resides entirely in $H_1$.)

It is now possible to apply (18) and (19) to compute the contribution of the two point
function to the $\zeta$ function associated with $\ln det^{-1/4} H_I^2$ and
$\ln det H_{II}$. We find that
$$\left. W^{(1)}(A) = -\frac{d}{ds}\right|_0\frac{1}{\Gamma(s)} \int_0^\infty
dt\,t^{s-1} tr\left\lbrace -\frac{1}{4} \exp - t \left(
\begin{array}{cc}
-D^2\delta_{ij} & -\epsilon_{imn}F_{mn}\\
\epsilon_{mnj}F_{mn} & -D^2\end{array}\right) +\exp - t(-D^2)\right\rbrace .\eqno(20)$$
Employing the expansion of eq. (19) and computing functional traces in momentum space with [20]
$$<p|f|q> = \frac{1}{(2\pi)^{3/2}}\, f(p-q)\eqno(21)$$
we find that the contribution to the two point function coming
from (20) is
$$\left. W^{(1)}_2(A) = -\frac{d}{ds}\right|_0\frac{1}{\Gamma(s)} \int_0^\infty
dt\,t^{s-1} \int \frac{dp\,dq}{(2\pi)^3}\left\lbrace
-\frac{1}{4} (- t)^2 \int^1_0 du\right. \eqno(22)$$
$$\left. e^{-\left[(1-u)p^2 + uq^2\right]t}
\left[\left(-\epsilon_{imn}\right)\left(f^{apb} f_{mn}^p (p-q)\cos
\frac{p \times q}{2}\right.\right.\right.\nonumber$$
$$\left. +d^{apb} f_{mn}^p (p-q)\sin\frac{p\times q}{2}\right)
\left(\epsilon_{rsi}\right) \left(f^{bqa} f_{rs}^q (q-p)\cos \frac{q \times p}{2}
\right.\nonumber$$
$$\left.\left.\left. +d^{bqa} f_{rs}^q (q-p) \sin \frac{q \times p}{2}\right)\right]\right\rbrace
\;\;\;\left(f_{ij}^a \equiv \partial_i A_j^a - \partial_j A_i^a\right).\nonumber$$
Upon making the usual shift in momentum variables, this becomes
$$\left. = \frac{N}{2} \frac{d}{ds}\right|_0 \frac{1}{\Gamma(s)} \int_0^\infty
dt\, t^{s+1} \int \frac{dp \,dq}{(2\pi)^3} \int_0^1 du\,e^{-\left[q^2 + u(1-u)p^2\right]t}
\nonumber$$
$$f_{mn}^a(p)f_{mn}^b (-p)\left(\delta^{ab} - \delta^{a0}\delta^{b0} \cos(p \times q)
\right).\eqno(23)$$
The standard integrals [21]
$$\int \frac{d^n k}{(2\pi)^n} e^{-k^2t} = \frac{1}{(4\pi t)^{n/2}}\eqno(24a)$$
$$\int_0^\infty dt\,t^{\nu - 1}e ^{-At} = \Gamma(\nu)A^{-\nu}\eqno(24b)$$
$$\int_0^\infty dt\,t^{\nu - 1} e^{-\gamma t - \beta /t} = 2\left(
\frac{\beta}{\gamma}\right)^{\nu / 2} K_{\pm \nu} (2\sqrt{\beta\gamma})\eqno(24c)$$
can be used to reduce (23) to
$$\left. W_2^{(1)} (A) = \frac{N}{2} \frac{d}{ds}\right|_0 \frac{1}{\Gamma(s)}
\int \frac{d^3 p}{(4\pi)^{3/2}} \int_0^1 du\,f_{mn}^a (p) f_{mn}^a (-p)\eqno(25)$$
$$\left[\Gamma\left(s + \sbody12 \right)\left(u(1-u)p^2\right)^{-s-1/2}\delta^{ab}\right.
\nonumber$$
$$\left.\left. -2 \left(\frac{\tilde{p}^2}{4u(1-u)p^2}\right)^{\frac{s+1/2}{2}} 
K_{s+1/2}\left(2
\sqrt{\frac{u(1-u)p^2\tilde{p}^2}{4}}\right)\delta^{a0}\delta^{b0}\right)\right]\nonumber$$
where $p \times q \equiv \tilde{p} \cdot q$.  It is now possible to evaluate
$\displaystyle{\left. \frac{d}{ds}\right|_0}$ in (25) explicitly, leaving us with
$$W_2^{(1)} (A) = 
\frac{N}{2} \int \frac{d^3 p}{(4\pi)^{3/2}} 
\int_0^1 du\,f_{mn}^a (p) f_{mn}^b (-p)
\left[\sqrt{\frac{\pi}{u(1-u)p^2}}\,\delta^{ab}\right.\eqno(26)$$
$$\left. -\left(\frac{4\tilde{p}^2}{u(1-u)p^2}\right)^{1/4} K_{1/2}
\left(\sqrt{u(1-u)p^2\tilde{p}^2}\right)\delta^{a0}\delta^{b0}\right].\nonumber$$
Since $K_{\pm 1/2}(z) = \sqrt{\frac{\pi}{2z}}e^{-z}$, it is possible to compute
both integrals over $u$ in (26) using [21]
$$\int_0^1 du\,u^{a-1}(1-u)^{b-1} = \frac{\Gamma(a)\Gamma(b)}{\Gamma(a+b)}\eqno(27a)$$
$$\int_0^1 du \frac{1}{\sqrt{u(1-u)}} e^{-A\sqrt{u(1-u)}} = 2 \int_0^{\pi /2} 
d\theta e^{-\frac{A}{2}\cos\theta} = \pi I_0 (A/2)\eqno(27b)$$
leaving us with
$$W_2^{(1)} (A) = \frac{N}{16}\int \frac{d^3p}{\sqrt{p^2}} \left[f_{mn}^a
(p) f_{mn}^b (-p)\left(\delta^{ab} - I_0\left(\frac{\sqrt{p^2\tilde{p}^2}}{2}\right)
\delta^{a0}\delta^{b0}\right)\right].\eqno(28)$$
In the limit $\theta^{\mu\nu} \rightarrow 0$ (so that $\tilde{p}^ 2 
\rightarrow 0$),
$I_0 \left(\frac{\sqrt{p^2\tilde{p}^2}}{2}\right) \rightarrow 1$ leaving only the $SU(N)$
contribution to (28).

We can now consider the loss of phase associated with the replacement of eq. (12).
This phase has been discussed extensively in [1, 10-12]; it is associated with
so-called $\eta$-function,
$$\eta(s) = \frac{1}{\Gamma\left(\frac{s+1}{2}\right)} \int_0^\infty dt\,t^{\frac{s-1}{2}}
Tr\left(H_I e^{-H_I^2t}\right). \eqno(29)$$
If a parameter $\lambda$ is inserted into $H_I$ so that
$H_I(\lambda = 1) = H_I$, then from (29) it is easily seen that
$$\frac{d\eta_\lambda (s)}{d\lambda} = \frac{-s}{\Gamma\left(\frac{s+1}{2}\right)}
\int_0^\infty dt\,t^{\frac{s-1}{2}} Tr\left(\frac{dH_I(\lambda)}{d\lambda} 
e^{-H_I^2(\lambda)t}\right).\eqno(30)$$
As only $\eta(0)$ is required to determine the phase we are 
interested in, it is sufficient to compute the poles arising in the integral over
$t$ in (30) on account of the explicit factor of $s$ arising in front of the
integral.

If now
$$H_I (\lambda) \equiv \left(
\begin{array}{cc}
\epsilon_{ipj}\left(\partial_p - i\lambda [A_p\right) & 
- \left(\partial_i -i\lambda [A_i\right)\\
\left(\partial_j - i\lambda [A_j\right) & 0\end{array}\right)\eqno(31)$$
then it is possible to expand the right side of (30) in powers of the 
background field using a second expansion due to Schwinger [20]
$$e^{-\left(H_0 + H_1\right)t} = \left[ e^{-H_0t} + (-t) \int_0^1
du\,e^{-(1-u)H_0t} H_1 e^{-u H_0 t}\right.\nonumber$$
$$+ (-t)^2 \int_0^1 du\,u \int_0^1 dv\,e^{-(1-u)H_0 t} H_1 e^{-u(1-v)H_0t}\nonumber$$
$$\left.  H_1 e^{-uvH_0t} + \ldots \right].\eqno(32)$$
From (15), (31) and (32), it is evident that the contribution
to (30) that is bilinear in the background field $A_\mu$ is
(upon following the approach used above with the $\zeta$-function)
$$\frac{d\eta_\lambda^{(2)} (s)}{d\lambda} = \frac{N\lambda s}{\Gamma\left(
\frac{s+1}{2}\right)} \int_0^\infty dt\,t^{\frac{s+1}{2}} \int_0^1 du \int
\frac{dp\,dq}{(2\pi)^3} e^{-\left(q^2 + u(1-u)p^2\right)t}\eqno(33)$$
$$\left(\epsilon_{ijk} A_i^a(p) f_{jk}^b(-p)\right)
\left(\delta^{ab} - \delta^{a0}\delta^{b0} \cos (p \times q)\right).\nonumber$$
The integrals appearing in (33) are identical in form to these in (23);
just as we obtain (28) we find that
$$\frac{d\eta_\lambda^{(2)} (s)}{d\lambda} = \frac{N\lambda s}{\Gamma\left(
\frac{s+1}{2}\right)} \int_0^1 du \int \frac{dp}{(4\pi)^{3/2}} \epsilon_{ijk}
A_i^a(p) f_{jk}^b (-p)\nonumber$$
$$\left[\delta^{ab}\Gamma\left(\frac{s}{2}\right)\left(u(1-u)p^2\right)^{-s/2}
\right.\eqno(34)$$
$$\left. -2 \delta^{a0}\delta^{b0}\left( \frac{\tilde{p}^2}{4u(1-u)p^2}\right)^{s/4}
K_{s/2}\left(\sqrt{u(1-u)p^2\tilde{p}^2}\right)\right].\nonumber$$
In the limit $x \rightarrow 0$, $K_\nu (x) \rightarrow 2^{\nu - 1} \Gamma(\nu)
x^{-\nu}$ and thus when $\theta_{\mu\nu} \rightarrow 0 (\tilde{p}^2 \rightarrow 0)$,
(34) reduces to
$$\approx \frac{N\lambda s}{\Gamma\left(\frac{s+1}{2}\right)} \int_0^1 du\int
\frac{dp}{(4\pi)^{3/2}}\,\epsilon_{ijk} A_i^a (p) f_{jk}^b (-p)\nonumber$$
$$\left[ \Gamma \left(\frac{s}{2}\right)\left(u(1-u)p^2\right)^{-s/2}\right]\nonumber$$
$$\left[\delta^{ab} - \delta^{a0} \delta^{b0}\right].\eqno(35)$$
As in the case of the $\zeta$-function, only the $SU(N)$ contribution
 to $\frac{d}{d\lambda} \eta_\lambda (0)$ survives in the
commutative limit.

\section{Discussion}

Chern Simons theory is difficult to regulate on account of the presence of the
tensor $\epsilon_{ijk}$. Operator regularization appears however to be
a suitable way of dealing with this problem as it does not involve altering the
original Lagrangian.  This permits one to deal with both the usual
commutative $SU(N)$ Chern Simons model and as well noncommutative $U(N)$ Chern
Simons theory. An analysis of the one loop two point function done here has shown that
in the commutative limit, the non-commutative $U(N)$ model smoothly reduces to the
commutative $SU(N)$ model.

\section{Acknowledgements}

Must of this work was done while the author
was a guest at the University of Sao Paulo; he is
particularly grateful to F. Brandt and J. Frenkel
for their hospitality there.  R. and D. MacKenzie had
useful suggestions.

\eject

\eject


\begin{thebibliography}{99}
\bibitem{1} E. Witten, Comm. Math. Phys. 121, 351 (1989).
\bibitem{2} D. Birmingham, M. Blau, M. Rakowski and G. Thompson,
Phys. Rep. 209, 129 (1991).
\bibitem{3} SenHu, ``Chern-Simons-Witten Theory'' (World Scientific, Singapore, 2002).
\bibitem{4} R. Szabo, hep-th 0109162.
\bibitem{5} M. Douglas and N.A. Nekrasov, hep-th 0106048.
\bibitem{6} G.H. Chen and Y.S. Wu, Nucl. Phys. B593, 562 (2001)\\
N. Grandi and G.A. Silva, Phys. Lett. B507, 345 (2001)\\
M.M. Sheikh-Jabbari, Phys. Lett. B510, 247 (2001)\\
A. Das and M.M. Sheikh-Jabbari, JHEP 0106, 28 (2001)\\
C.P. Martin, Phys. Lett. B515, 185 (2001).
\bibitem{7} A.S. Bichl, J.M. Grimstrup, V. Putz and M. Schweda, JHEP 0007, 046 (2001).
\bibitem{8} D.G.C. McKeon and T.N. Sherry, Phys. Rev. Lett. 59,
532 (1987), Phys. Rev. D35, 854 (1987).
\bibitem{9} A. Salam and J. Strathdee, Nucl. Phys. B90, 203 (1975)\\
J. Dawker and R. Critchley, Phys. Rev. D13, 3224 (1976)\\
S. Hawking, Comm. Math. Phys. 55, 133 (1977).
\bibitem{10} P. Gilkey, ``Invariance Theory, the Heat Equation,
and the Atiyah-Singer Index Theorem'', Publish or Perish Inc. (1984).
\bibitem{11} D. Birmingham, R. Kantowski and M. Rakowski, Phys. Lett. B
151, 121 (1990); Phys. Rev. D4, 3476 (1990).
\bibitem{12} D.G.C. McKeon and T.N. Sherry, Ann. of Phys. 218, 325 (1992).
\bibitem{13} D.G.C. McKeon, Can. J. Phys. 68, 1291 (1990).
\bibitem{14} D.G.C. McKeon and C. Wong, Int. J. of Mod. Phys. A10, 2181 (1995).
\bibitem{15} F.A. Dilkes, L.C. Martin, D.G.C. McKeon and T.N. Sherry, Int. J. of
Mod. Phys. A14, 463 (1999).
\bibitem{16} B.S. DeWitt, Phys. Rev. 162, 1195, 1239 (1967).
\bibitem{17} G. 't Hooft, Nucl. Phys. B61, 455 (1973).
\bibitem{18} L. Abbott, Nucl. Phys. B185, 189 (1981).
\bibitem{19} L. Bonara and M. Salizzoni, Phys. Lett. B504, 80 (2001).
\bibitem{20} J. Schwinger, Phys. Rev. 82, 664 (1951).
\bibitem{21} I. Gradshteyn and M. Ryzhik, Table of Integrals,
Series and Products (Academic Press, New York 1980).
\end{thebibliography}
\end{document}